\documentclass[preprint, 3p]{elsarticle}

\usepackage{graphics}
\usepackage{subcaption}
\usepackage{rotating}
\usepackage{bm}
\usepackage{mathrsfs}
\usepackage{amssymb,amsmath}
\usepackage{float}
\usepackage{siunitx}
\usepackage{blindtext}
\usepackage{mathtools}
\usepackage{xcolor}
\usepackage{csquotes}
\usepackage{tikz}
\usepackage{float}
\usepackage{textcomp}
\usepackage{soul} 

\usepackage[colorlinks,linkcolor=red,citecolor=blue,urlcolor=blue]{hyperref}

\biboptions{sort&compress} 

\usepackage{amssymb}
\usepackage{pifont}
\newcommand{\cmark}{\ding{51}}%
\newcommand{\xmark}{\ding{55}}%

\journal{Journal of Computational Science}

\newcommand{\feq}{f^\mathrm{eq}}
\newcommand{\fneq}{f^\mathrm{neq}}

\newcommand{\fone}{f^{(1)}}

\newcommand{\uu}{\bm{u}}
\newcommand{\uj}{\bm{\xi}}
\newcommand{\uN}{\bm{\nabla}}
\newcommand{\uQ}{\bm{Q}}
\newcommand{\uS}{\bm{S}}

\newcommand{\uI}{\bm{I}}
\renewcommand{\O}{\mathcal{O}}
\newcommand{\p}{\partial}
\newcommand{\KN}{\mathrm{Kn}}
\newcommand{\MA}{\mathrm{Ma}}
\newcommand{\RE}{\mathrm{Re}}

\newcommand{\ux}{\bm{x}}
\newcommand{\uxi}{\bm{\xi}}

\title{Local mesh refinement sensor for the lattice Boltzmann method}

\author[cui]{Y. Thorimbert\corref{cor1}}
\ead{yann.thorimbert@unige.ch}

\author[nz]{D. Lagrava}

\author[orestis]{O. Malaspinas} 

\author[cui]{B. Chopard}

\author[cui]{C. Coreixas}

\author[cui]{J. de Santana Neto}

\author[ralf]{R. Deiterding}

\author[cui]{J. Latt}
\ead{jonas.latt@unige.ch}

\address[cui]{Department of Computer Science, University of Geneva, 1227 Carouge, Switzerland}

\address[orestis]{HEPIA - Haute école du paysage, d'ingénierie et d'architecture, 1202 Geneva, Switzerland}

\address[nz]{University of Canterbury, Civil and Natural Resources Engineering, 8041 Christchurch, New Zealand}

\address[ralf]{School of Engineering, University of Southampton, Boldrewood Campus, SO16 7QF, United Kingdom}

\cortext[cor1]{Corresponding author}

\date{\today}

\begin{document}

\begin{abstract}

A novel mesh refinement sensor is proposed for lattice Boltzmann methods (LBMs) applicable to either static or dynamic mesh refinement algorithms. The sensor exploits the kinetic nature of LBMs by evaluating the departure of distribution functions from their local equilibrium state. 
This sensor is first compared, in a qualitative manner, to three state-of-the-art sensors: (1) the vorticity norm, (2) the Q-criterion, and (3) spatial derivatives of the vorticity.
This comparison shows that our kinetic sensor is the most adequate candidate to propose tailored mesh structures across a wide range of physical phenomena: incompressible, compressible subsonic/supersonic single phase, and weakly compressible multiphase flows.
As a more quantitative validation, the sensor is then used to produce the computational mesh for two existing open-source LB solvers based on inhomogeneous, block-structured meshes with static and dynamic refinement algorithms, implemented in the Palabos and AMROC-LBM software, respectively.
The sensor is first used to generate a static mesh to simulate the turbulent 3D lid-driven cavity flow using Palabos. AMROC-LBM is then adopted to confirm the ability of our sensor to dynamically adapt the mesh to reach the steady state of the 2D lid-driven cavity flow. Both configurations show that our sensor successfully produces meshes of high quality and allows to save computational time. 
\end{abstract}


\maketitle

\section{Introduction}

The lattice Boltzmann method (LBM) is widely used in the field of computational fluid dynamics, in which it has proven its importance with respect to other traditional numerical methods in multiple fields, including incompressible, weakly and fully compressible flows as well as complex multiphysics applications~\cite{doi:10.1146/annurev.fluid.30.1.329,bi:chopard_cellular_2005,GUO_Book_2013,HUANG_Book_2015,KrugerBook,SUCCI_Book_2018}.
In all kinds of mesh-based approaches to computational fluid dynamics, a local adaptation of the size of mesh cells is of importance. Indeed, to save on computational expense, applications often require to resolve a wide range of physical scales (see~\cite{Parmigiani2013,labchip} for instance). In other cases, such as in external aerodynamics and aeroacoustics simulations, the computational domain may be much larger than the actual fluid domain of interest, to limit the effect of artificial domain boundaries on the flow structure~\cite{COREIXAS_PEGASUS_AIAA_2015,SENGISSEN_AIAA_2993_2015,BROGI_JASA_142_2017,APPELBAUM_AIAA_2973_2018,FEUCHTER_CF_224_2021}. In all these cases, simulations benefit substantially from a local increase of spatial resolution in critical regions of the computational domain, which are most often unknown in advance. There exists therefore a need for sensors capable of identifying areas in which the mesh requires local adjustments.

Mesh refinement is in itself a highly complex topic in the field of LB methods, which are traditionally designed to be executed on homogeneous meshes with uniform mesh density. 
In a canonical LB scheme based on a collision-streaming sequence, the streaming step copies data to neighboring cells at relative positions provided by lattice constants which are immutable and as such constrain the mesh to a uniform shape. For a local adaptation of the mesh size, it is common practice to use structured meshes with cubic elements which are organized in a hierarchic structure such as an octree, in which the relative edge length of neighboring cells typically varies by powers of two. In this way, a normal LB scheme can be executed within groups of equally sized cells, and an additional mesh refinement algorithm is carried out to transfer data across refinement levels (see, e.g.,~\cite{filippova_grid_1998,dupuis_theory_2003,chen_grid_2006,RHODE_IJNMF_51_2006,lagrava_revisiting_2012,GENDRE_PRE_96_2017,ASTOUL_PhD_2021}). Depending on the relative spatial arrangement of cells at different levels, refinement algorithms are split into vertex-centered and cell-centered schemes, which are both presented in Section~\ref{sec:mesh-refinement-overview}.

Most works dedicated to grid refinement algorithms focus on how to adapt the mesh instead of providing a way to predict regions where mesh refinement is required. 
In addition, these works usually rely on static meshes obtained through \emph{a priori} knowledge of the flow features, such as boundary layer development and separation for wall-bounded flows. 
In contrast, mesh refinement sensors and error estimators predict, with minimal or no \emph{a priori} knowledge, areas where refinement is needed. 
By coupling these sensors with adaptive mesh refinement techniques, one then drastically reduces user inputs for efficient and accurate simulations.
In that context,
Crouse et al.~\cite{crouse_lb-based_2003} propose a sensor based on the velocity divergence as an error indicator in an incompressible flow. Eitel-Amor et al.~\cite{Eitelamor2013} use an approach based on two sensors, one proportional to the velocity norm and the other providing a measure of the total pressure. Fakhari \& Lee~\cite{fakhari2014} compare different methods applied to a finite-difference LB scheme. The comparison includes the Q-criterion, which is proportional to the symmetric and the anti-symmetric part of the velocity-derivative tensor. Other interesting error indicators are the vorticity norm and the derivatives of the vorticity norm in specific directions. 
The latter criterion is used in the context of wall-bounded turbulent flows which require an accurate simulation of boundary layers.
Beyond the literature on LBM, other error indicators can be found in the literature for finite-volume Navier-Stokes solver with block-structured AMR (SAMR). An overview is proposed in Kamkar et al.~\cite{kamkar_feature-driven_2011}. It includes the Q-criterion and so-called lambda-sensors which, based on eigenvalues of velocity-derivative tensors, provide an estimate for the presence of vortical or other coherent structures. 

In the present work, a novel sensor is proposed, which is computed from the populations available on a given cell and is therefore fully local. It detects the need for further mesh refinement based on an expectation of how the ratio between off-equilibrium and equilibrium parts of the velocity distribution functions converges towards the global Knudsen number ($\KN$). In convective scaling, the Knudsen number is simply a constant, independent of the level of grid refinement. As discussed in Section~\ref{sec:algorithm}, this sensor has a clear physical interpretation: deviations of the locally computed Knudsen number from its theoretical value must be compensated by a finer resolution.

The article is structured as follows: The theoretical part first provides a summary of the theory of mesh refinement algorithms (Section~\ref{sec:mesh-refinement-overview}). Although the choice of a mesh refinement sensor in principle depends on the physics of the considered problem rather than on the specifics of a given mesh refinement algorithm, this summary provides additional context to the numerical verification part, in which two codes with different mesh refinement strategies are used to validate the new sensor. The theoretical part then introduces some concepts and notations of the LBM used in the present study. The verification in Section~\ref{sec_numerics} starts with a qualitative comparison of the proposed sensor against other sensors proposed in the literature as a way to predict its usefulness for different types of fluid problems. Then, simulation results are presented in which the criterion is used to build a static mesh for a 3D cavity flow in Palabos (with vertex-centered grid refinement) and a dynamic mesh for a 2D cavity flow in the SAMR code AMROC-LBM (with cell-centered grid refinement). The quality of the produced results is discussed in both cases.

\section{Theory}
\subsection{Mesh refinement}\label{sec:mesh-refinement-overview}
Lattice Boltzmann refinement algorithms can be split into vertex-centered schemes~\cite{filippova_grid_1998,dupuis_theory_2003,lagrava_advances_2012,yu_multi-block_2006,geller_benchmark_2006,pellerin_implementation_2015,GENDRE_PRE_96_2017,ASTOUL_JCP_418_2020,ASTOUL_JCP_447_2021}, in which some of the coarse and fine nodes are situated at coinciding coordinates along the refinement interface, and cell-centered schemes~\cite{RHODE_IJNMF_51_2006,yu_interaction_2009,Eitelamor2013,SCHORNBAUM_SIAM_38_2016}, in which coarse and fine meshes have a staggered relative arrangement. The latter category is compatible with the point of view adopted by the community of SAMR~\cite{berger_local_1989}, a field traditionally dedicated to finite volume codes that has in recent years been applied to LB schemes~\cite{dillmann_adaptive_2016,deiterding_predictive_2016,feldhusen2016,grondeau_2021,barad_lattice_2017}. Another important distinguishing feature between mesh refinement algorithms is the choice of the value for the time step $\delta t$ in mesh levels with different cell edge length $\delta x$. In diffusive scaling, the ratio $\delta x/\delta t^2$ remains constant across mesh levels, thus guaranteeing asymptotic mesh convergence for incompressible flow problems, while in acoustic scaling, which is naturally adapted to compressible flows, the ratio $\delta x/\delta t$ is constant. Many authors however adopt acoustic scaling for both compressible and incompressible flows, because only this choice can guarantee the continuity of both density and pressure across the interface of a refined mesh~\cite{lagrava_advances_2012}. In addition, the diffusive scaling usually leads to very small time steps, as the number of refinement levels increases, because moving from coarse to finer levels decreases the time step by a factor four. As a last remark, it is worth noting that the various proposed algorithms further differ in other aspects, including the presence or absence of time interpolation schemes.

In this work, our refinement sensor is used to generate the mesh in the two LB codes Palabos~\cite{LATT_CMA_81_2021} and AMROC-LBM~\cite{dillmann_adaptive_2016,feldhusen2016,grondeau_2021}. Palabos adopts the vertex-centered-based mesh-refinement algorithm described in Ref.~\cite{lagrava_advances_2012}, which is based on: (1) third-order polynomial interpolation, (2) coarse-to-fine filtering, and (3) acoustic scaling. This algorithm has been re-used and further improved in recent publications~\cite{BROGI_JASA_142_2017,GENDRE_PRE_96_2017,ASTOUL_JCP_418_2020,ASTOUL_JCP_447_2021}. 
In the AMROC framework~\cite{amroc}, dynamic structured mesh adaptation is implemented generically on finite volume meshes and with the same recursive algorithm~\cite{berger_local_1989} that makes use of ghost cells to prescribe boundary conditions and achieve communication between refinement blocks. The LB variant AMROC-LBM uses a cell-centered scheme, in which streaming is performed first, followed by collision. Acoustic scaling is used to couple levels; interface populations propagating across refinement boundary are meticulously tracked, cf.~\cite{feldhusen2016,dillmann_adaptive_2016} for details. Palabos and AMROC-LBM are used to generate the data compiled in Section~\ref{sec:grid-refinement}.

\subsection{Lattice Boltzmann method}\label{sec_lbm}

LBM is by now a well-known numerical method in computational fluid dynamics. This section presents only basic concepts, while the reader is referred to Refs.~\cite{doi:10.1146/annurev.fluid.30.1.329,bi:chopard_cellular_2005,GUO_Book_2013,HUANG_Book_2015,KrugerBook,SUCCI_Book_2018} for further details.

The Boltzmann equation (BE) describes the time evolution of 
large numbers of particles in a region of the space $\ux \in \mathbb{R}^3$ with a given velocity 
$\uj \in \mathbb{R}^3$, which are represented by the particle mass distribution function $f(\ux,\uj,t)$. 
The Boltzmann equation for a gas without external force reads
\begin{equation}\label{eq:boltzmannEq}
 \p_t f + \uj\cdot \uN f = -\Omega(f,\feq),
\end{equation}
with the collision operator $\Omega$ and where $\feq$ is the equilibrium distribution function, given by the Maxwell-Boltzmann distribution. For the rest of this discussion, the BGK collision operator is used: $\Omega = \frac{1}{\tau} (f - \feq)$, with $\tau$ a relaxation time~\cite{BHATNAGAR_PR_94_1954}.

Following the ideas presented in the works by Shan et al.~\cite{SHAN_PRL_80_1998,SHAN_JFM_550_2006}, one can discretize the velocity space to a set of $q$ velocities $\bm\xi_i$, leading to velocity-discrete populations $f_i(\ux,t) \equiv f(\ux,\uj_i,t)$, $i\in [0,q-1]$. In this work, a D2Q9 and a D3Q27 models have been used.

As will be seen later, it is convenient to reformulate Eq.~\eqref{eq:boltzmannEq} as a function of non-dimensional quantities; from now on in this section, variables expressed in physical units will be written with a star for the sake of clarity (\textit{e.g.} the physical macroscopic velocity $u^*$), whereas non-dimensional variables are written normally. The reformulation is done using characteristic time $t_0^*$, length $l_0^*$, density $\rho_0^*$ and velocity magnitude $c_0^*$ (the isothermal speed of sound). The velocity-discrete version of Eq.~\eqref{eq:boltzmannEq} with the BGK operator then reads:
\begin{equation}\label{eq:nondimDVLBE}
        \frac{1}{t_0^*}\partial_{ t}  f_i + \frac{c_0^*}{l_0^*}\bm{\xi}_i\cdot \uN  f_i = -\frac{1}{\tau^*}\left(  f_i -  f_i^{\textrm{eq}}\right),
\end{equation}
where $ f_i = f_i^* / \rho_0^*$ are the non-dimensional populations and $\bm {\xi}_i = \bm\xi_i^* / c_0^*$ are the non-dimensional mesoscopic velocities. Similarly, non-dimensional operators have been used : $\partial_t=\partial^*_t t^*_0$ and $\uN=\uN^* l^*_0$. The condition $l_0^*=c_0^*t_0^*$ is required in order to represent the correct macroscopic behaviour. Eq.~\eqref{eq:nondimDVLBE} will be used below in order to relate the Knudsen number of the flow to the particle distribution functions.

A non-dimensionalization and a time and space discretization (with time step $\delta t^*$ and spacing $\delta x^*$) lead to the LBM equation that is actually solved~\cite{DELLAR_CMA_65_2013}:
\begin{equation}\label{eq:bgklb}
\bar f_i(\ux+ \uj_i \delta t^* / t_0^*,t+1) - \bar f_i(\ux,t) = -\frac{1}{\tau} (\bar f_i (\ux,t)- \bar f_i^{\mathrm{eq}}(\ux,t)),
\end{equation}
where $\tau=\tau^*/\delta t^* + 1/2$, and
\begin{equation}\label{eq:lbm}
    \bar f_i =  f_i + \frac{\delta t^*}{2\tau^*} \left( f_i -  f_i^{\mathrm{eq}} \right).
\end{equation}
Note that the condition $l_0^*=c_0^*t_0^*$ mentioned above, together with the geometric requirement $| \uj_i| \delta t^* / t_0^* = \delta x^*/l_0^*$ for straight velocities, implies that $\delta x^*/\delta t^*=c_0^*/c_s$, where $c_s=1/\sqrt 3$ represents the magnitude of the straight velocities. Finally, the choices $\delta t^*=t_0^*$ and $\delta x^*=l_0^*/c_s$ simplify the form of Eq.~\eqref{eq:lbm} as chosen for the implementation.

The discrete equilibrium distribution function is expressed by the truncated Maxwellian equilibrium~\cite{SHAN_PRL_80_1998,SHAN_JFM_550_2006}
\begin{equation}\label{eq:feq_i}
    \bar f_i^{\mathrm{eq}}= f_i^{\textrm{eq}} = w_i\rho\left(1+\frac{\uxi_i\cdot\uu}{c_s^2}+\frac{1}{2c_s^4}\uQ_i:\uu\uu\right),
\end{equation}
where $\rho$ is the density, $\uu$ is the macroscopic velocity field, $\uQ_i=\uxi_i\uxi_i-c_s^2\uI$, and $w_i$ the lattice weights. From now on, the bar on the populations is omitted.
The density and the velocity fields are computed by the distribution function through the relations
\begin{align}
\rho&=\sum_{i=0}^{q-1}f_i=\sum_{i=0}^{q-1}\feq_i,\label{eq:rho}\\
\rho\uu&=\sum_{i=0}^{q-1}\uxi_if_i=\sum_{i=0}^{q-1}\uxi_i\feq_i\label{eq:u}.
\end{align}
For implementation purposes, a time-step is decomposed into two parts that are applied successively on the whole 
computational domain. The two steps are called the ``collide-and-stream'' operation.
\begin{enumerate}
  \item The collision, which \emph{locally} modifies the value of the populations
  according to
  \begin{equation}
    f_i^\mathrm{out}(\ux,t)=f_i(\ux,t)-\Omega_i(f_i,\feq_i).
  \end{equation}
  \item The streaming, which moves the populations to their neighbors according to their microscopic velocity
  \begin{equation}
  f_i(\ux+\uxi_i,t+1)=f_i^\mathrm{out}(\ux,t).
  \end{equation}
\end{enumerate}

Performing a multi-scale Chapman-Enskog (CE) expansion
(see~\cite{bi:chapman60,bi:chopard_cellular_2005} for more details), one can show
that the LBM BGK scheme is asymptotically equivalent to the weakly compressible
Navier-Stokes equations
\begin{align}
\p_t\rho+\uN\cdot(\rho\uu)&=0, \label{eq:ns1}\\ 
\p_t(\rho\uu)+\uN\cdot(\rho\uu\uu)&=- \uN p+\uN\cdot(2\nu\uS), \label{eq:ns2}
\end{align}
with $p$ being the pressure, $\uS$ the viscous stress tensor and $\nu$ the kinematic viscosity defined by
\begin{align}
p&=c_s^2\rho,\label{eq:rho_p}\\
\uS&=\frac{1}{2}\left[\uN\uu+(\uN\uu)^\mathrm{T}\right],\\
\nu&=c_s^2(\tau-1/2).\label{eq:nu_omega}
\end{align}
From now on, the kinematic viscosity $\nu$ will be treated as a constant quantity.

The CE expansion is done under the assumption that $f_i$ can be expanded with respect to a small parameter $\varepsilon$
\begin{equation}\label{eq:decompositionF}
f_i=f_i^{(0)}+\varepsilon\fone_i+\O(\varepsilon^2),
\end{equation}
where $\varepsilon\ll 1$ can further be identified with the Knudsen number~\cite{bi:huang87}.
To obtain this result, one first uses Eq.~\eqref{eq:nondimDVLBE}, keeping only the lowest orders after replacing $f_i$ by the CE \emph{Ansatz}~\eqref{eq:decompositionF}. At $\mathcal{O}(1)$ and $\mathcal{O}(\varepsilon)$, one ends up with~\cite{SHAN_JFM_550_2006}
\begin{equation}\label{eq:generalf0}
f_i^{(0)} = f_i^{(eq)}(\rho, \uu, T),
\end{equation}
and
\begin{equation}\label{eq:generalf1}
\fone_i = -\tau \left[ \partial_t f_i^{(0)} + \uj\cdot \uN f_i^{(0)}\right]
\end{equation}
respectively.
From this, the non-equilibrium contribution $\fone_i$ naturally depends on time and space derivatives of all macroscopic quantities (density $\rho$, velocity $\uu$ and temperature $T$) through the zero-order contribution $f_i^{(0)}$. One can then expect $\fone_i$ to become non-negligible for strong, local variations of density, velocity or even temperature fields, whether they be of physical or numerical nature.

Finally, by further noticing that the derivatives should scale like the characteristic quantities of the system, the above relationship~(\ref{eq:generalf1}) becomes
\begin{equation}
 f_i^{(1)}  \sim  \tau^* \left(\frac{1}{t_0^*}+\frac{c_0^*}{l_0^*}\right)  f_i^{(0)}.
\end{equation}
By interpreting $\lambda^*=c_0^* \tau^*$ as the mean free path of particles, and using $c_0^*=l_0^*/t_0^*$, the Knudsen number $\KN = \lambda^* / l_0^*$ is introduced, yielding
\begin{equation} \label{eq:fNEqfEqRatio}
 \frac{ f_i^{(1)}}{  f_i^{(0)}} \sim \KN.
\end{equation}
Hence, this ratio should remain small in the continuum limit for smooth flows. On the contrary, one should expect it to become large close to discontinuities and multiphase/multicomponent interfaces (e.g., shock wave and liquid/gas interface respectively). We propose in \ref{deriv:dtf} an interpretation of this quantity and its use as a sensor in the case of incompressible flows.






\section{Refinement sensor} \label{sec:algorithm}

\subsection{Principles}
In this section, the refinement sensor is proposed independently of the actual algorithm used to implement grid refinement. Also, the stars are from now on omitted for variables expressed in physical units.

The proposed sensor is based on the fact that the off-equilibrium and equilibrium 
parts of the distribution function are linked to each other through the Knudsen number as defined above:
\begin{equation}
 \fneq \sim \feq \KN.
\end{equation}




Consider now a grid $G_c$ with mesh spacing $\delta x_c$, and a finer grid $G_f$ with mesh spacing $\delta x_\mathit{f}$ linked by the relationship $\delta x_\mathit{f} = \delta x_c / n$, where $n$ is a positive integer.
It is assumed that the same simulation is executed on both grids and that the Reynolds number $\RE$ is the same in both cases (the lattice relaxation time is adapted to ensure this constraint). Also, the ratio $\delta x / \delta t$ is set to be constant in $G_c$ and $G_f$, hence
$\delta t_c/\delta t_\mathit{f}=\delta x_c/\delta x_\mathit{f}$. With this constraint, called \enquote{convective scaling} or \enquote{acoustic scaling}, quantities that are proportional to the velocity are invariant in lattice units, and the Mach number does not depend on the grid resolution. Acoustic scaling however yields a proper convergence rate for compressible flow only and fails to approximate incompressible flow solutions with second-order convergence rate, as Eqs.~(\ref{eq:ns1}-\ref{eq:ns2}) that are recovered with the CE expansion contain error terms of order $\mathcal{O}(\delta t^2) + \mathcal{O}(\KN^2)$.




\subsection{Implementation}
To formalize the computation of the sensor, the following algorithm is proposed:
\begin{enumerate}
 \item Choose a uniform spatial resolution $\delta x$.
 \item Run the simulation.
 \item For all cells, compute the local Knudsen number averaged over all discrete velocities
\begin{equation}\label{eq:sensor}
C = \frac{1}{q} \sum_{i=0}^{q-1} \left| \frac{\fneq_i}{\feq_i} \right|.
\end{equation}
\end{enumerate}


It can be noted that the idea of using the ratio between off-equilibrium population (to a given power $n\in \mathbb{N}$) and their equilibrium counterpart was recently propose to locally stabilize simulations through local entropic filtering~\cite{GORBAN_PA_414_2014} and artificial viscosity~\cite{LATT_RSTA_378_2020,Coreixas2020} with $n=2$ and $1$ respectively.




\subsection{Comparison with other sensors}
 As discussed in the introduction, the proposed sensor will be compared to different other sensors in Section~\ref{sec_numerics}.
 The first one is the \enquote{vorticity norm sensor}:

\begin{equation}\label{eq:vorticity}
K_{\omega} =|\bm\omega|=|\uN \times \uu|,
\end{equation}
and the second one is based on the incompressible formulation of the \enquote{Q-criterion}~\cite{KOLAR_AIAAJ_53_2015}:
\begin{equation}
    K_{\textrm{Q}} =\frac{1}{2}\left( |\bm\Omega|^2 - |\bm S|^2 \right),
\end{equation}
with $\Omega= \frac{1}{2}\left[\uN\uu-(\uN\uu)^\mathrm{T}\right]$ the rotation rate tensor. Then, a sensor proposed by Fakhari \& Lee in~\cite{fakhari2014} is also tested here:
\begin{equation}\label{eq:fakhari}
    K_{\textrm{F}}=\frac{\sqrt{\left(\partial_x \omega\right)^2 + \left(\partial_y \omega\right)^2}}{|\bm\omega|}.
\end{equation}


Note that this latter sensor has been proven to yield accurate results in some cases, in particular for the 2D lid-driven cavity benchmark.
Finally, the proposed \enquote{Knudsen sensor} reads
\begin{equation} \label{eq:kkn}
     K_{\KN} = \frac{1}{q} \sum_{i=0}^{q-1} \left| \frac{\fneq_i}{\feq_i} \right|.
\end{equation}

\subsection{Practical considerations for the use of a refinement sensor}
Whereas the quantity $K$ for all refinement sensors listed in the previous section denotes refinement factors (\textit{i.e.} $\delta x_c = K\cdot \delta x_\mathit{f}$), it is convenient to work with the logarithm of $K$,
\begin{equation}
    \phi = \log_2 K,
\end{equation}
which is linked linearly with the desired refinement level $l$:
\begin{equation}
    l = 
    \begin{cases}
    l_{min} & \text{if $[a \cdot \phi] < l_{min}$}\\
    l_{max} & \text{if $[a \cdot \phi] > l_{max}$}\\
    [a \cdot \phi] & \text{otherwise},
    \end{cases}
\end{equation}
where the bracket $[x]$ denotes the integral part of a value $x$, and $l_{min}$ and $l_{max}$ stand for the minimal (coarsest) and maximal (finest) refinement level respectively. The scale factor $a$ is chosen empirically (with our sensor as well as with other sensors) to fit the values of $\phi$ encountered in the simulation to the desired range of refinement levels. In this way, the different degrees of refinement correspond to powers of 2, as required by the implementation:
\begin{equation}
    \delta x = 2^{l_{max}-l} \delta x_\mathit{f},
\end{equation}
where $\delta x_\mathit{f}$ denotes the size of the finest cells.

In practice, the accuracy and efficiency of a code using static or dynamic mesh refinement depends therefore not only on the choice of the refinement sensor, but also on a fine tuning of these scale parameters. This manuscript is not focusing on the technical aspects of this exercise in parameter tuning, nor on the relative accuracy obtained by different refinement sensors, which all need to be tuned individually for optimal results. Instead, Section~\ref{sec_numerics} provides a qualitative comparison of different sensors and discusses their capability to capture specific physical phenomena that are usually considered to require mesh refinement. In Section~\ref{sec:grid-refinement}, where the Knudsen sensor is used to implement static grid-refinement in a 3D LB simulation and AMR in a 2D LB simulation, the scale parameter $a$ is fixed choosing the finest refinement level $\delta_{x\mathit{f}}$ and the overall number of levels.

\section{Qualitative comparison of sensors for different types of physics}\label{sec_numerics}

Hereafter, we compare the ability of each sensor to highlight under-resolved regions for a wide range of physical phenomena: flow past a cylinder, lid-driven cavity flow, Riemann problem and droplet at rest. By further comparing these results with \emph{a priori} knowledge about each configuration, it is possible to identify which sensor is more likely to provide meaningful information to grid-refinement algorithms.

 \subsection{2D lid-driven cavity}\label{sec:cavity2d}
Wall-bounded flows are commonly encountered in the context of internal fluid dynamics. For these, it is of paramount importance to correctly capture the development of the boundary layers, and their interaction with the bulk flow. 
To investigate the ability of each sensor to properly identify these flow features, we propose to simulate a 2D lid-driven cavity flow for different Reynolds numbers: $\RE=1000$ and $3200$.
It consists of a square domain resolved by $N \times N$ lattice cells. A no-slip boundary condition $\uu = 0$ is applied on each wall except the top wall located at $y=N$, where the condition $\uu = (u_{LB},0)^T$ is applied. The moving bounce-back condition as described in~\cite{KrugerBook} was used on the walls to impose the velocity boundary conditions. In order to perform the tests at Reynolds number $\RE=10^3$, a Smagorinsky subgrid-scale model~\cite{Smagorinsky} is added to the collision step, with Smagorinsky parameter $c_\mathrm{smago}=0.14$.

A simulation with parameters $\RE=1000$, $u_{LB}=5\cdot 10^{-2}$ and $N=120$ has been performed in order to compare the different sensors. For this benchmark, the sensor of Fakhari \& Lee $K_F$ is chosen as the reference sensor, following the conclusions of~\cite{fakhari2014}, which show that this sensor is highly accurate for the cavity problem at $\RE=3200$.

\begin{figure*}[!tbp]
\centering
\includegraphics[trim={0 0.5cm 0 0},clip,width=\textwidth]{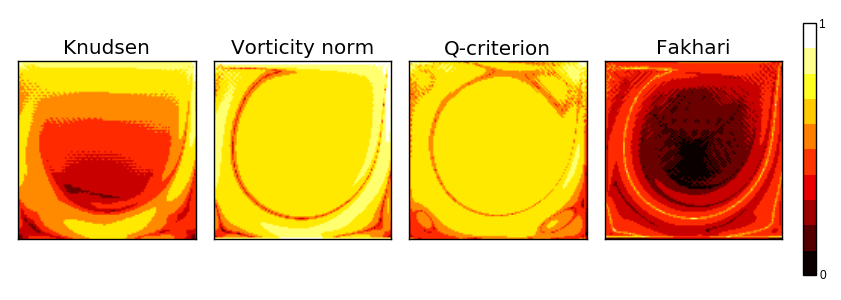}
\caption{Average refinement level in the case of the 2D lid-driven cavity with $\RE = 10^3$,  $u_{LB}=5\cdot 10^{-2}$ and $N=120$.}
\label{fig:compCav2D}
\end{figure*}

Figure~\ref{fig:compCav2D} displays the normalized average refinement level $\phi$ in the whole simulation domain and for each sensor. It is found that the Knudsen sensor is similar to the one of Fakhari \& Lee at a global scale, as both suggest a low resolution in the center of the domain. Close to the borders on the other hand, the values proposed by the Knudsen sensor are similar to those of the vorticity-norm sensor. 

While the sensor of Fakhari \& Lee, which was designed for this purpose, may produce a mesh somewhat better adapted to the needs of the boundary layer, the Knudsen sensor once again suggests overall reasonable mesh levels matching the expectations qualitatively.

\subsection{Flow past a 2D cylinder}\label{sec:cylinder2d}
As a first benchmark test, we investigate how sensors identify physical phenomena related to the simulation of a flow past bluff bodies, e.g., boundary layer development and wake formation. More precisely, we consider the flow past a cylinder with a simulation domain that consists of a $n_x \times n_y$ rectangle with a bottom wall at $y=0$ and a top wall at $y=n_y$. A cylinder of radius $R=n_y/10$ is centered in $x_c=n_x/3$. To trigger the unsteadiness of the flow, the $y$-coordinate of the cylinder center is slightly offset from the centerline and set to $y_c = 9n_y/20$.  Simulations have been performed with $n_x=300$, and for $\RE = 100$ and $\RE = 1000$. In all cases, $n_y=n_x/2$. A no-slip boundary condition is applied to the cylinder surface and the lateral walls, using the same bounce-back method as in the 2D cavity benchmark described above. A Zou/He velocity boundary condition~\cite{zou_pressure_1997} is used at the inlet in $x=0$ to impose a velocity $u_{LB}$ chosen to match the target Reynolds number. The outlet consists of a Neumann boundary condition : $f_i(t,x=n_x)$ is set to $f_i(t,x=n_x-1)$ for every population $i$ and for every time step $t$. A Smagorinsky subgrid model~\cite{Smagorinsky} was adopted to perform the simulations, with Smagorinsky parameter $c_\mathrm{smago}=0.14$. For this benchmark, the vorticity norm is chosen as the reference sensor, as suggested in~\cite{fakhari2014} for the case of a the flow past a square cylinder.

\begin{figure*}[!tbp]
\centering
\includegraphics[trim={0 0.3cm 0 0},clip,width=\textwidth]{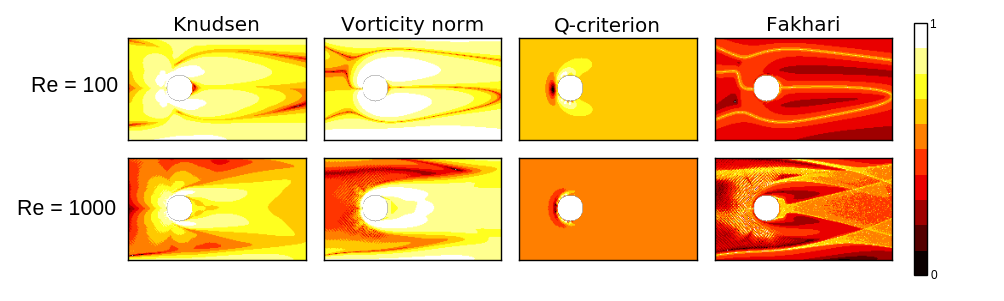}
\caption{Comparison of the sensors for the flow around a 2D cylinder at (top) $\RE=1000$ and (bottom) $\RE=1000$. The normalized $\phi$ value for each sensor is shown in the subdomain $x\in[x_c-4R,x_c+10R]$, $y\in[y_c-4R,y_c+4R]$.}
\label{fig:cylinder_re_all}
\end{figure*}

Figure~\ref{fig:cylinder_re_all} depicts the time average value of $\phi$ for each sensor at $\RE=100$ and $\RE=1000$. It is observed that the Knudsen sensor is essentially similar to the vorticity norm in this case, for $\RE=100$ as well as for $\RE=1000$. The spatial average of the refinement factor is comparable for the Knudsen sensor and the vorticity norm at high Reynolds number, whereas the vorticity norm yields lower values at low Reynolds number (see Tab.~\ref{tab:spaceAvg}). 

\subsection{3D flow past a sphere}\label{sec:sphere3d}
In this test the domain size is $n_x=400$, $n_y=160$ and $n_z=160$. In this problem, the domain is periodic along $y$-axis and $z$-axis. The inlet is situated at $x=0$, and an outlet sponge zone, in which the relaxation parameter is linearly increased, spreads from $x=300$ to $x=400$. The velocity $\bm u(x=0,y,z)=(u_{LB},0,0)^T$ at the inlet ($Y-Z$ plane at $x=0$) is imposed through a regularized boundary condition~\cite{Latt2008}, with $u_{LB}=3\cdot 10^{-2}$. A sphere of diameter $D=20$ lattice nodes, modelled through a bounce-back boundary condition, is located at $x_c=100$, $y_c=80$, $z_c=80$. The simulation is executed with a BGK collision model and Smagorinsky subgrid-scale modeling~\cite{Smagorinsky} (Smagorinsky parameter $c_\mathrm{smago}=0.14$). As in the 3D cavity test, the Palabos library~\cite{LATT_CMA_81_2021} was used. The results were averaged after transient state was reached, from iteration $4\cdot 10^4$ to iteration $5\cdot10^4$. Just like for the corresponding 2D benchmark case, the vorticity norm serves as a reference sensor, because it is known to properly single out the area around the sphere and vortex-shedding phenomena in its wake.

\begin{figure}[H]
\centering
\includegraphics[width=1.\textwidth]{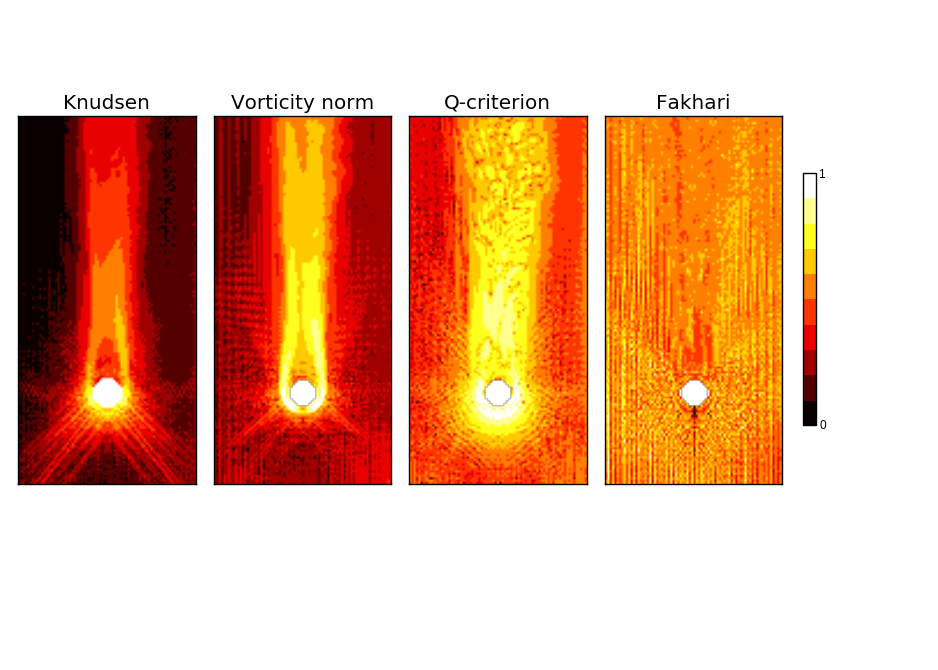}
\caption{Comparison of time-averaged refinement sensors for a flow around a sphere at $\RE=1887$, at a deliberately poor resolution. The normalized $\phi$ value for each sensor is shown in the subdomain $x\in[0.4n_x,0.8n_x]$, $y\in[0.25n_y,0.75n_y]$ and $z=0.5n_z$.}
\label{fig:sphereHigh}
\end{figure}
\begin{figure}[H]
\centering
\includegraphics[width=1.\textwidth]{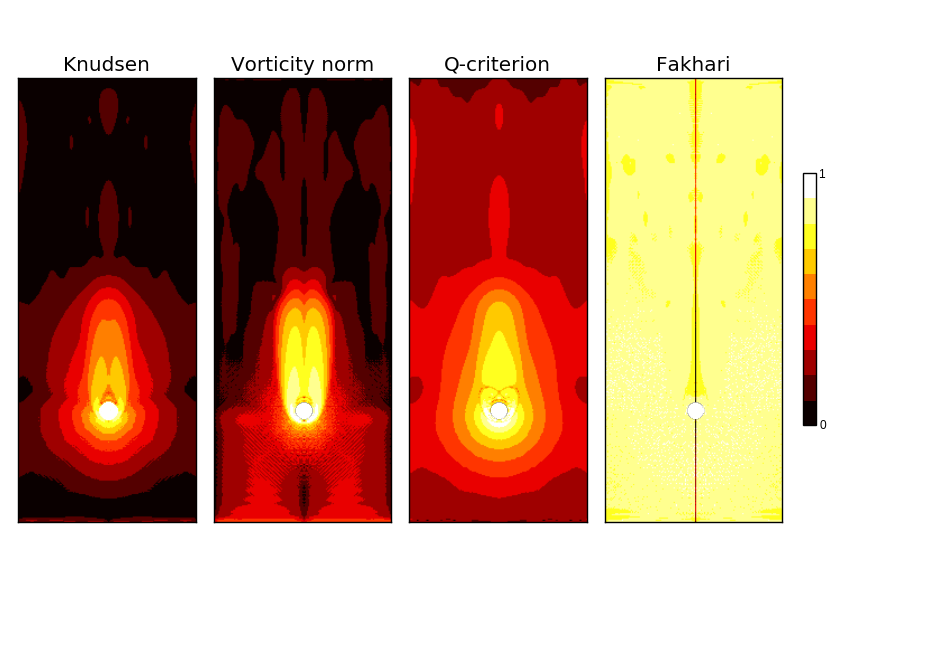}
\caption{Comparison of the time average of the refinement sensor for the flow around a sphere at $\RE=75$.}
\label{fig:sphereLow}
\end{figure}

The normalized value of $\phi$ for each sensor is displayed in Figure~\ref{fig:sphereHigh} for $\RE=1887$ and Figure~\ref{fig:sphereLow} for $\RE=75$. Both the Knudsen sensor and the Q-criterion follow the lead of the vorticity and properly predict the requirement for high mesh resolution around the obstacle and in its wake. Solely, the Q sensor appears less useful at high Reynolds numbers, as it pushes for highest-level mesh resolution in large swaths of the full computational domain. The sensor of Fakhari \& Lee finally does not exhibit useful values, neither at low nor at high Reynolds number, as expected for a sensor that is mainly targeted at interior flow problems.

The resolution of the simulations is kept intentionally low to test the capability of a sensor to predict the need for further resolution while running in an underresolved regime. In this case, spurious numerical patterns occur in the area between the obstacle and the inflow, which are first visible in the pressure field and velocity gradients (on Figure~\ref{fig:sphereHigh}, they are most visible in the Knudsen and the vorticity-norm sensors). They can be traced back to the boundary representation of the obstacle, here a low-resolution bounce-back scheme and can lead to numerical instabilities. Clearly, the occurrence of such patterns calls for better mesh resolution. Thus, the Knudsen sensor, which singles out these patterns very neatly, turns out to be particularly useful in this respect.


 \subsection{2D Riemann problem}
 This 2D benchmark case of a compressible flow (see~\cite{Schulz1993} for a detailed description) enables the study of flows with sharp density, velocity and temperature variations. It is considered as a challenging test case due to the complex wave patterns arising in the solution~\cite{Kurganov2002}. Recently, Coreixas \& Latt proposed a shock capturing technique that locally adds artificial viscosity in regions highlighted by the Knudsen sensor. This helped stabilizing the simulation while properly capturing the key flow features of this Riemann problem, in particular the shock wave fronts and their interplay~\cite{Coreixas2020}. As a consequence, the Knudsen sensor is assumed to provide reference data on this configuration.
 
 A square domain of $L\times L$ lattice nodes is set with $L=1000$, and the initial conditions described in Tab.~\ref{tab:shock}. The domain is periodic along $X-$axis and $Y-$axis avoid the need for time-dependent boundary conditions. Owing to the compressible nature of this problem, the methodology described in Ref~\cite{LATT_RSTA_378_2020} has been used to run the simulation. More precisely, the D2Q49 lattice was used along with a BGK collision model based on a numerical equilibrium, that enforces all 13 constraints required to recover the Navier-Stokes-Fourier equations in an \emph{exact manner}. The simulation corresponds to an inviscid flow ($\nu=0$), with a reference temperature $T_0=0.7$, and results are plotted for $(x,y)\in \left[L/4,3L/4\right]^2$ at iteration number 150.
 
 \begin{table*}[!tbp]
\centering
\begin{center}
\begin{tabular}{ l|c|c|c|c } 
 ~ & quadrant 1 & quadrant 2 & quadrant 3 & quadrant 4 \\
 \hline
 $\rho$ & 0.513 & 1 & 0.8 & 1 \  \\
  $u_x$ & 0 & 0.7276 & 0 & 0 \  \\
 $u_y$ & 0 & 0 & 0 & 0.7276 \  \\
 $P$ & 0.4 & 1 & 1 & 1 \  \\
\end{tabular}\caption{Initial conditions of the 2D Riemann problem, with the pressure denoted by $P$. In the visual representations of the simulation, quadrant 1 is located on the upper right corner of the domain, quadrant 2 in the upper left corner, quadrant 3 in the bottom left corner and quadrant 4 in the bottom right corner.}
\label{tab:shock}
\end{center}
\end{table*}
 
\begin{figure*}[!tbp]
\centering
\includegraphics[trim={0 0.5cm 0 0},clip,width=\textwidth]{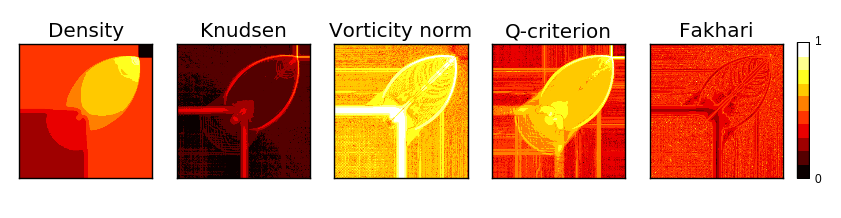}
\caption{Comparison of the sensors for the Riemann problem. The normalized $\phi$ value for each sensor is shown in the center of the simulation domain ($x,y\in [0.25,0.75]$), along with $log(\rho)$ for comparison.}
\label{fig:shock}
\end{figure*}

Figure~\ref{fig:shock} displays on the left-most image the normalized logarithm of the density $\rho$ to help identify regions that naturally require high grid resolution due to sharp density variations. The four remaining images show the $\phi$ value of the four sensors, again represented by a normalized version $\phi_n$ that is rescaled to the $[0, 1]$ range. Only the center part of the domain is shown ($x,y\in [0.25,0.75]$). By visual assessment, it is clearly determined that vorticity norm and the Fakhari \& Lee sensors do not properly capture shock wave fronts propagating in quadrant 1 toward the upper right corner, as opposed to the Knudsen sensor and the Q-criterion. Moreover, the Knudsen sensor appears better adapted to the needs of mesh refinement than the Q-criterion, as it calls for finer meshes in smaller, more sharply depicted areas of the domain. This is confirmed by computing the spatial average of each sensor: $\bar{\phi}_{n,\textrm{Kn}}=0.14$, whereas $\bar{\phi}_{n,\textrm{Q}}=0.49$. For a more in-depth discussion of the characteristics of this flow problem, the reader is referred to~\cite{Coreixas2020}.
  
It is interesting to emphasize that in this benchmark case, the Knudsen sensor proves to be very different, and substantially more useful than the vorticity-norm sensor, although the two sensors produced rather similar results in previous single phase incompressible flow cases. 
The Knudsen sensor proves more versatile, by virtue of the gradients of different macroscopic variables present in the non-equilibrium population (Eq.~\ref{eq:fneq}). While velocity gradients are dominant in single phase incompressible flows, leading to an expression akin to the one of the vorticity norm, all gradients of macroscopic quantities (density, velocity and temperature) should be properly captured for this compressible example. 

\subsection{2D droplet}\label{sec:droplet}
Hereafter, a weakly compressible multiphase flow is simulated to understand how sensors react in the presence of non-negligible (1) density gradients and (2) spurious currents. More precisely, this multiphase benchmark consists in a 2D droplet which is at rest in a periodic square domain of size $N\times N$, with a resolution $N=200$ cells.
The droplet is simulated as a circular region of the domain centered about $(x_c,y_c) = (N/2,N/2)$, with a radius $R=N/4=50$ cells. The liquid density is set to $1.95$, whereas the rest of the domain is initialized with a gas density equal to $0.15$, eventually leading to a density ratio of $13$. Lattice populations are initialized at equilibrium with the aforementioned density and with zero velocity. The simulation is stopped after a stationary state is reached. A Shan-Chen single component model~\cite{shan1993lattice} with parameter $G=-5$ was used to simulate the phases of the problem. The pseudo-potential force was accounted for through Guo's forcing methodology~\cite{Guo2002} along the BGK collision model.

\begin{figure*}[!htbp]
\centering
\includegraphics[width=1.\textwidth]{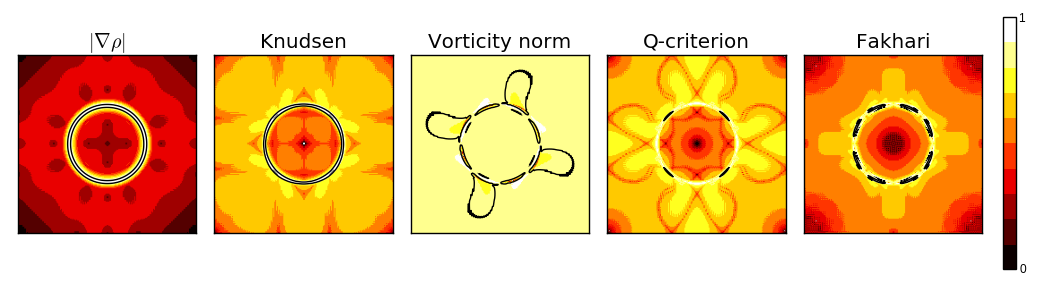}
\caption{Normalized refinement level yielded by the different sensors in the case of the 2D droplet in the stationary state using a single component Shan-Chen model. The norm of the density gradient is used as a reference in this simulation. The contour line (black) corresponds to a value of 0.45}
\label{fig:droplet}
\end{figure*}

Figure~\ref{fig:droplet} shows the refinement levels predicted by each sensor at the end of the simulation, along with the density gradient for reference. It is observed that the Knudsen sensor captures the interface accurately, while suggesting a low resolution inside the droplet and isotropy issues due to spurious currents, as the Q-criterion and the Fakhari \& Lee sensor do. However, the contour lines demonstrate that the Knudsen sensor puts a more consistent focus on the interface of the droplet than other criteria. This can be explained by the presence of non-negligible spurious currents which more severely impact Fakhari \& Lee, Q-criterion, and the vorticity norm sensors, with the latter one being of little interest for such a configuration. To isolate the impact of parasitic currents on the above conclusion, one could rely on the well-balanced LB formulation recently proposed by Guo~\cite{GUO_PoF_33_2021}. The latter allows for the drastic reduction of the amplitude of spurious currents, even with the simple BGK collision model. Such an investigation is left for future work. 

\section{Use of the Knudsen sensor as a grid refinement sensor}\label{sec:grid-refinement}
This section aims at confirming the accuracy and efficiency improvement induced by the use of our kinetic based refinement criterion. With this idea in mind, we first rely on this criterion to build a static mesh density field that is further used to simulate the 3D lid-driven cavity flow at $\mathrm{Re}=12000$ on a non-uniform mesh in an efficient and accurate manner. Then, it is shown that such an approach can be adopted in the 2D context for adaptive mesh refinement. In the latter case, the mesh density is (re)computed in a dynamic manner to speedup the convergence toward steady state.

\subsection{3D cavity at $\RE=12000$ with static mesh refinement}\label{subsec:3DcavityPalabos}
In this first test simulation with grid refinement, the Knudsen sensor is used to generate a mesh for a 3D lid-driven cavity at a Reynolds number $\RE=12000$. The accuracy of the results is verified by comparison against a Chebyshev spectral simulation reported in~\cite{Leriche2006}. To set up this problem, a no-slip condition was imposed on each wall, except on the lid ($x$-$z$ plane at $y=h$) where an $x$-velocity was imposed with a smooth velocity transition on the lid edges, as described in~\cite{Leriche2000}:

\begin{equation}
u(x,y=h,z) = U_0\left[1 - \left(\frac{x}{h}-1\right)^{18}\right]^2 \left[1 - \left(\frac{z}{h}-1\right)^{18}\right]^2,
\end{equation}
where $U_0$ is the maximum lid velocity. Here, $h=1$ and $U_0$ and the lattice relaxation parameter is chosen in order to match the Reynolds number For $\RE=12000$, $\MA\approx 0.14$ and $\KN\approx 1.15\cdot 10^{-5}$.

\begin{figure*}[!htb]
\centering
\includegraphics[trim={0 0.5cm 0 0},clip,scale=.25]{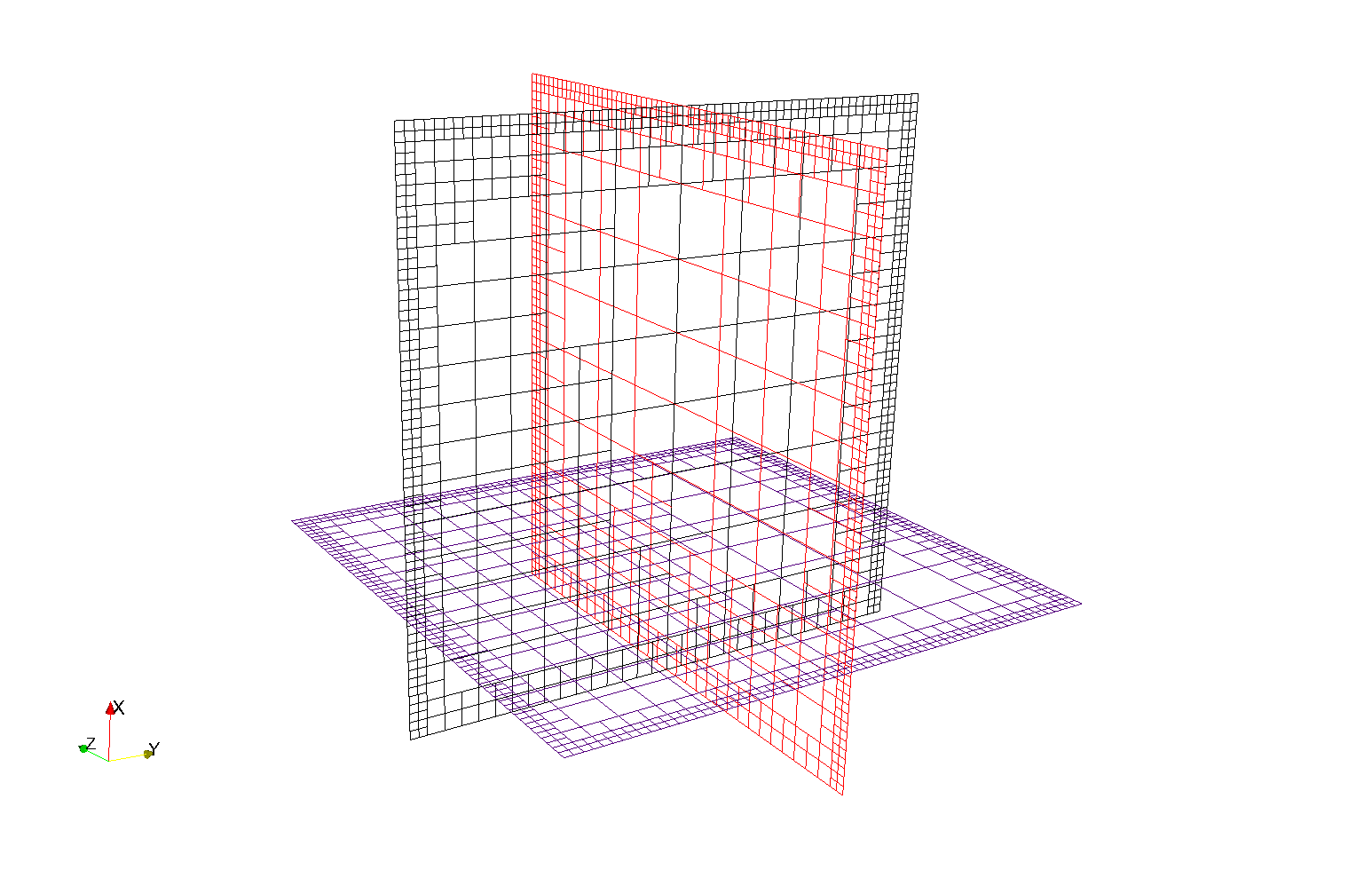}
\caption{Schematic representation of the generated grid for the 3D cavity. For the sake of clarity, cell sizes were increased by a factor 16 on this picture.}
\label{fig:grid3D}
\end{figure*}

A simulation with $200\times 200\times 200$ lattice nodes has first been performed on a uniform, relatively coarse mesh in order to compute the Knudsen sensor and build a corresponding, refined mesh. For this purpose, the refinement sensor was averaged over the duration of one dimensionless time period starting at the end of the initial, transient flow. 
Along the walls, the grid was set uniformly to the highest resolution to prevent the boundary conditions from crossing refinement interfaces. 
In the rest of the simulation domain, the resolution can vary from $N=64$ to $N=512$ depending on the local value yielded by the sensor. The generated grid is represented in Figure~\ref{fig:grid3D} with cells that are coarser than the actual computational cells, for the sake of visualization. In total, the generated grid contains approximately 11 million cells, whereas a uniform grid at the finest mesh resolution would contain approximately 134 million cells.

\begin{figure*}[!htb]
\centering
\includegraphics[scale=.45]{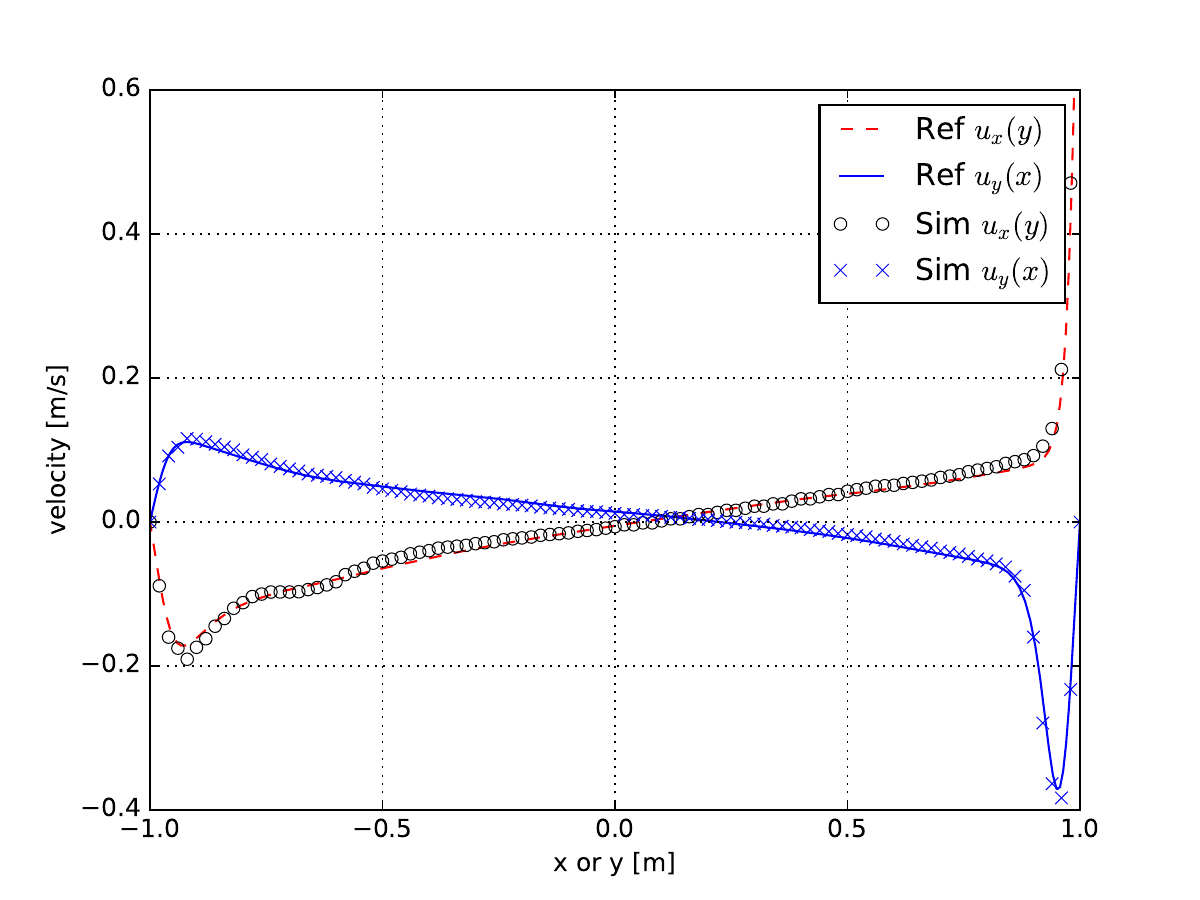}
\caption{Comparison of the averaged centerline $u_x$ and $u_y$ velocity for the 3D cavity, for the cases of the reference solution and the static refined grid.}
\label{fig:uCavity3D}
\end{figure*}

Figure~\ref{fig:uCavity3D} shows a comparison of the mean $u_x$ and $u_y$ velocities in the reference solution and in the LB simulation with the refined grid. The velocity values from the LB simulation are averaged from $t=300$ to $t=450$ dimensionless time cycles. This long simulation time is required to gather sufficient statistically data to reach convergence, and it is one of the reasons for which this benchmark case is known to be very challenging. The comparison shows an excellent match and as such demonstrates the adequacy of the mesh used to run the problem.

\subsection{2D cavity with adaptive mesh refinement}\label{subsec:amr}
In this second practical use case of the Knudsen sensor, a dynamic, time-dependent mesh is generated for the 2D cavity problem as described in Section~\ref{sec:cavity2d}, adopting the AMROC-LBM software~\cite{dillmann_adaptive_2016,deiterding_predictive_2016,feldhusen2016}. The employed LB scheme is the BGK operator in two space dimensions, as described in Section~\ref{sec_lbm}. The AMROC-LBM results here use linear temporal interpolation, second-order accurate spatial interpolation and re-scaling of the non-equilibrium part $\fneq$ of populations at the interface between levels,  cf.~\cite{grondeau_2021}, as originally proposed by Dupuis \& Chopard~\cite{dupuis_theory_2003}. 

The simulation is performed at a Reynolds number of $\RE=1000$ with a multi-level structured AMR grid, based on a coarsest grid resolution with $N=50$ cells across the width of the cavity. While several large eddy simulation models have been implemented for AMROC-LBM~\cite{gkoudesnes_2021}, the present results do not apply any such model. Although AMROC allows arbitrary refinement factors, 2 is used here for all additional refinement levels. Evaluation of the Knudsen criterion and mesh reorganization is carried out at the beginning of every level-0 time step. One additional layer of buffer cells is marked for refinement around cells tagged by the criterion. Refinement blocks are created with an efficiency of 80\%, meaning that at least 80\% of cells in each block need to be flagged. A block is further sub-divided otherwise. Proper nesting is also enforced, ensuring that a refinement is fully contained in the next coarser level with at least one row of coarse buffer cells. The detailed topological algorithms in AMROC can be found in \cite{amroc}. 

\begin{figure*}[!htb]
\centering
\includegraphics[width=1.\linewidth]{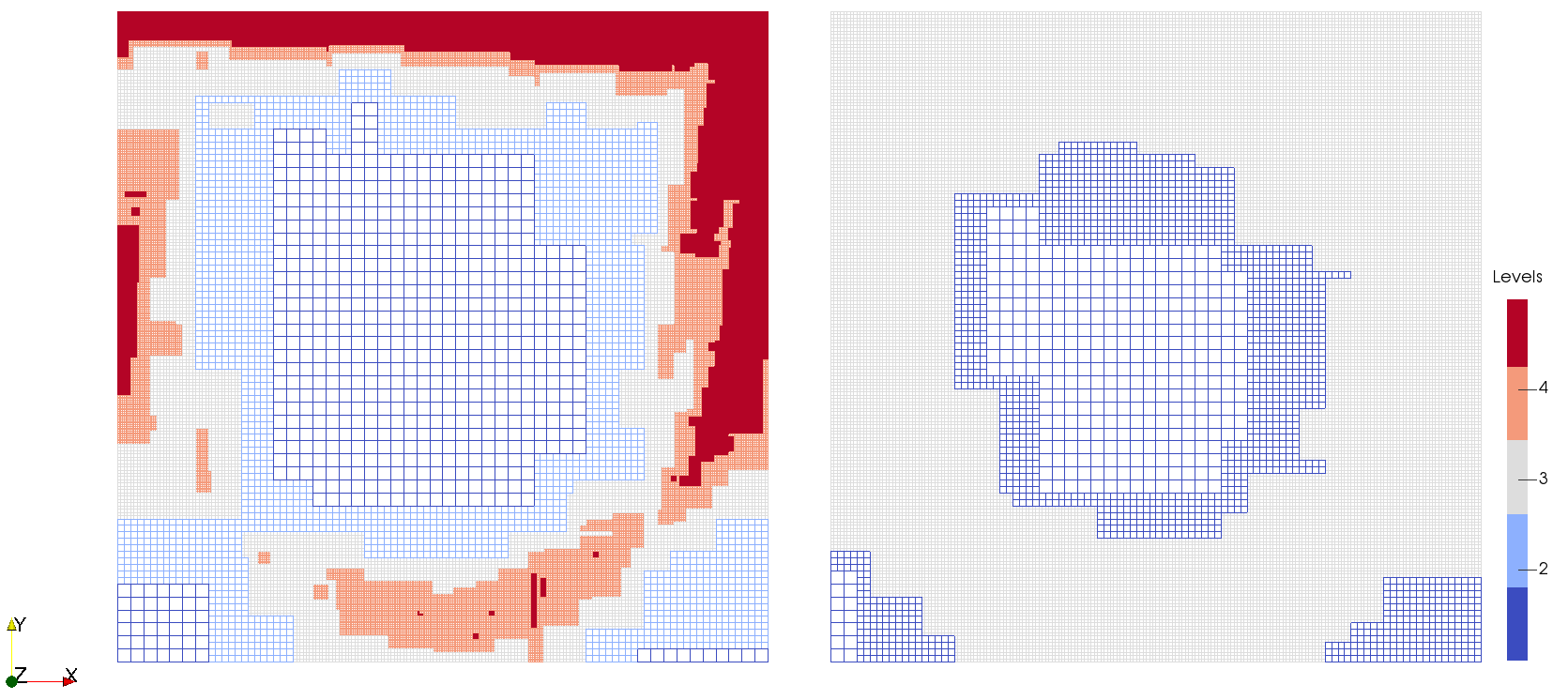}
\caption{Snapshot of 2D cavity simulations using adaptive mesh refinement with (left) five and (right) three refinement levels.}
\label{fig:amr}
\end{figure*}

Figure~\ref{fig:amr} shows a snapshot of the grid generated by AMROC-LBM, based on the Knudsen sensor, during the simulation, for both 3 levels and 5 levels of refinement. A video of the simulation is provided as supplementary material to this article.

For quantitative comparison, the velocity profiles of $u_x$ along the $Y$-axis centerline and of $u_y$ along the $X$-axis centerline were evaluated after convergence of the flow and compared to reference values presented in~\cite{Ghia_Ghia_Shin_1982}. Table~\ref{tab:amr} lists the root-mean-square error between the computed velocities and the reference values along the two profiles. The number of cells for the refined mesh in Tab.~\ref{tab:amr} is computed in terms of effective cells, weighted by their frequency of execution in the convective scaling regime:
\begin{equation}
    N_{\mathrm{effective}} = \sum_{i=1}^{L} \frac{n_i}{2^{L-i}},
\end{equation}
where $n_i$ stands for the total number of cells at level $i$ and $L$ the finest grid level. This comparison shows that AMR using our sensor allows accurate results while significantly saving the number of lattice points, as compared to a homogeneous grid made of cells the same size as the finest level of the refined grid.

\begin{table}
\centering
\begin{center}
\begin{tabular}{ l|c|c|c } 
 Simulation & RMSE $u_x$ & RMSE $u_y$ & $N_\mathrm{effective}$\\ 
 \hline
 Homogeneous mesh ($N=50$) & 0.060 & 0.053 & 2500 \  \\
 Homogeneous mesh ($N=200$) & 0.015 & 0.018 & 40000 \  \\
 Adaptive mesh using $K_\textrm{Kn}$ & 0.014 & 0.027 & 27672 \  \\
\end{tabular}\caption{Root-mean-square error of homogeneous mesh simulation and adaptive mesh simulation from the reference solution.}
\label{tab:amr}
\end{center}
\end{table}

\section{Discussion and conclusion}\label{sec_conclusion}

A novel, local grid refinement sensor has been proposed. It benefits from the kinetic nature of LBMs through the local evaluation of the departure of distribution functions from their equilibrium state, hence providing valuable information on under-resolved regions of the simulation domain for a wide range of physical phenomena. This sensor was compared to three state-of-the-art criteria over five benchmarks and compared qualitatively with expectations based on a physical understanding of the investigated flows. 


Table~\ref{tab:accuracy} summarizes the outcome of this qualitative discussion by assigning a grade to each of the sensor in every case. Furthermore, to support the discussion below, Tab.~\ref{tab:spaceAvg} summarizes the space average of the refinement factor obtained with each sensor on the different benchmarks, as a measure of the reduction of the number of grid cells that can be achieved by using the sensor. 

Our proposed sensor 
is found to produce either excellent or acceptably good results in all executed test cases. As shown in the summary in Tab.~\ref{tab:accuracy}, it is the only one among the four tested sensors that would produce an acceptable mesh across the full range of investigated physical situations. It therefore constitutes a sensor of choice for simulations in which too little \emph{a priori} knowledge of the problem is available to produce the computational mesh manually. Moreover, this method has the advantage to rely only on local information (i.e., the populations) and is therefore very efficient in terms of parallelism, as compared to sensors that require the computation of gradients of macroscopic quantities through finite differences or similar means. Since it needs no \emph{a priori} knowledge of the system, this method is also an excellent candidate for adaptive grid refinement techniques in time-dependent problems, as demonstrated by the preliminary results given in Section~\ref{subsec:amr}. 

For future work, we plan to apply the Knudsen sensor to produce static and AMR meshes for more advanced problems and assess its efficiency in comparison with other approaches. Another extension to the present work will also include an investigation of the possibility to combine the Knudsen sensor with another criterion capable of compensating for potential weaknesses of the Knudsen sensor for specific physical phenomena, such as the adaptation to a thin near-wall fluid layer.

\begin{table}
\centering
\begin{center}
\begin{tabular}{ l|c|c|c|c } 
 ~ & $K_{\textrm{Kn}}$ & $K_{\omega}$ & $K_{\textrm{Q}}$ & $K_{\textrm{F}}$\\
 \hline
 Cylinder (Re=100) & \cmark & ref. & \xmark & \xmark \  \\
 Cylinder (Re=1000) & \cmark & ref. & \xmark & \xmark \  \\
 Lid-driven Cavity & \texttildelow & \texttildelow & \texttildelow & ref. \  \\
 Riemann & \cmark & \xmark & \texttildelow & \xmark \  \\
 Droplet & \cmark & \xmark & \texttildelow & \cmark \  \\
\end{tabular}\caption{Summary of the accuracy of the sensors for each benchmark. A check mark (\cmark) indicates a good match with the reference sensor, a tilde (\texttildelow) indicates a poor though relevant match, and a cross mark (\xmark) indicates a poor match and the fact that the sensor cannot be used as a refinement sensor for this problem. Where a sensor has been used as the reference, it is indicated as \enquote{ref.} in the corresponding line.}
\label{tab:accuracy}
\end{center}
\end{table}

\begin{table}
\centering
\begin{center}
\begin{tabular}{ l|c|c|c|c } 
 ~ & $\bar K_{\textrm{Kn}}$ & $\bar K_{\omega}$ & $\bar K_{\textrm{Q}}$ & $\bar K_{\textrm{F}}$\\
 \hline
 Cylinder (Re=100) & 0.77 & 0.82 & 0.77 & 0.38 \  \\
 Cylinder (Re=1000) & 0.71 & 0.70 & 0.67 & 0.46 \  \\
 Lid-driven Cavity & 0.54 & 0.71 & 0.66 & 0.32 \  \\
 Riemann & 0.14 & 0.67 & 0.49 & 0.41 \  \\
 Droplet & 0.63 & 0.85 & 0.64 & 0.55 \  \\
\end{tabular}\caption{Summary of the space average of the refinement factor for each sensor on the different benchmarks. If the sensor is able to correctly identify areas of the simulation domain that require local refinement patches, then the lower the spatial average value, the better.}
\label{tab:spaceAvg}
\end{center}
\end{table}

\section*{Acknowledgment}
We acknowledge financial support from the Swiss National Science Foundation (SNSF) through project grants 200020\_197223 and 200021\_165984.

\appendix

\section{Derivation of $\delta t_\mathit{f}$ as a function of $\delta t_c$ for incompressible flows}\label{deriv:dtf}

To give the reader an insight of the relationship between coarse and fine grid parameters, we restrict the discussion to the incompressible case in this appendix.

The \textit{computed} Knudsen number in the coarse grid, noted $\KN_c$, does not correspond to the actual, physical $\KN$. In the other hand, the computed Knudsen number $\KN_f$ in the refined grid corresponds to the physical Knudsen number $\KN$ by definition, as one requires that the refined grid is such that it doesn't need to be refined anymore. Hence a term $\KN/\KN_c$ is expected to be included in the conversion factor between the grids.

We now want to express $\delta t_\mathit{f}$ as a function of $\delta t_c$, $\KN_c$ and $\KN$. The off-equilibrium population, in a given lattice, takes the form
\begin{equation}\label{eq:fneq}
    \fneq \sim    \left(\frac{\tau}{\delta t}+\frac{1}{2}\right) \delta t A,
\end{equation}
where $A$ denotes all the gradients on macroscopic quantities. Note that the operator $\nabla$ do not depend on the resolution.

Since the velocity is the same in both lattices as $\feq_c = \feq_f$, it follows that $\KN / \KN_C = \fneq_f / \fneq_c$. Hence,
\begin{equation}\label{eq:knratio}
    \frac{\KN}{\KN_c} = \frac{2\tau+\delta t_\mathit{f}}{2\tau+\delta t_c}.
\end{equation}


According to Eq. \eqref{eq:nu_omega}, and using $c_0=c_s \delta x / \delta t$, the physical viscosity can be expressed as
\begin{equation}\label{eq:visc}
    \nu =c_0^{2}\tau,
\end{equation}
which, using Eq.~\eqref{eq:knratio}, finally leads to

\begin{equation}
    \delta t_\mathit{f} = \delta t_c \frac{\KN}{\KN_c} + \frac{2\nu}{c_0^{2}}\left(\frac{\KN}{\KN_c}-1\right),
\end{equation}
where the second term will be neglected in the next sections. Indeed, since $\nu/c_0^2 = \tau$, the condition $\KN \ll 1$, under which is done the perturbation expansion of Eq.~\eqref{eq:decompositionF}, can be formulated as $c_0 \tau \ll l_0$, hence $\tau \ll \delta t$ for a given lattice.

\end{document}